\documentclass[a4paper,11pt]{article}
\pdfoutput=1 

\usepackage{jinstpub} 

\usepackage[british]{babel}

\usepackage{hyperref}

\usepackage{graphicx}
\usepackage{wrapfig}
\usepackage{caption}
\usepackage{color}
\usepackage{amsmath}
\usepackage{subfigure}
\usepackage{acro}
\usepackage{booktabs} 
\usepackage[squaren, Gray, cdot, named, prefixed, abbr]{SIunits}
\usepackage{bm}

\DeclareAcronym{OPA}{short={OpAmp},long=Operational Amplifier}
\DeclareAcronym{IC}{short=\textsc{IC},long=Ionization Chamber}
\DeclareAcronym{UHDR}{short=\textsc{UHDR},long=Ultra High Dose Rate}
\DeclareAcronym{FLASH-RT}{short=\textsc{FLASH-RT},long=FLASH Radiotherapy}
\DeclareAcronym{CONV-RT}{short=\textsc{CONV-RT},long=Conventional Radiotherapy}
\DeclareAcronym{BCT}{short=\textsc{BCT},long=Beam Current Transformer}
\DeclareAcronym{DosiFLASH}{short=\textsc{DosiFLASH},long=FLASH-RT Dosimeter}
\DeclareAcronym{ID}{short=\textsc{ID},long=Inner Diameter}
\DeclareAcronym{PSD}{short=\textsc{\textbf{PSD}},long=Power Spectral Density of noise}
\DeclareAcronym{RMS}{short=\textsc{RMS},long=Root Mean Square}
\DeclareAcronym{DR}{short=\textsc{DR},long= Droop Rate}
\DeclareAcronym{BLR}{short=\textsc{BLR},long= BaseLine Restoration}
\DeclareAcronym{GBP}{short=\textsc{GBP},long=Gain Band Product}
\DeclareAcronym{TI}{short=\textsc{TI},long=Trans-Impedance}
\DeclareAcronym{EF}{short=\textsc{EF},long=ElectronFlash machine}
\DeclareAcronym{EM}{short=\textsc{EM},long=Electro-Magnetic}

\title{Low noise optimization of an electron beam current transformer for conventional radiotherapy up to ultra high dose rate irradiations.}
\author[1]{C.~Lahaye\note{Corresponding author},}
\author{J.-M.~Fontbonne,}
\author[2]{S.~Salvador \note{\url{https://orcid.org/0000-0003-3444-7807}}}
\affiliation{Normandie Univ, ENSICAEN, UNICAEN, CNRS/IN2P3, LPC Caen, 14000 Caen, France }

\emailAdd{lahaye@lpccaen.in2p3.fr}

\abstract{
Ultra high dose rate electron beams also known as FLASH radiotherapy is becoming of importance in several preclinial cancer treatment studies. However, due to the dose rate used during the irradiation sessions, no real time dose monitoring device exists to date. In this work, we present the development of a beam current transformer (BCT), from the choice of the ferromagnetic component and the realisation of the shielding to the design of a front-end electronics based on a trans-impedance circuit in order to perform a low noise optimization of the detector. The BCT prototype is able to monitor a beam current range from 1.2~\textmu A to 200~mA with a rise time constant better than 20~ns and a droop rate of the signal below 0.05\%$\cdot$\textmu s$^{-1}$. Preliminary in-situ measurements are also presented. The goal is to combine the  BCT system which measure in real time the beam current, to an ionisation chamber monitoring the beam shape and position in order to provide a reliable dose monitoring system.}


\arxivnumber{2206.05948} 

\begin{document}
\bibliographystyle{unsrt}

\maketitle

\section{Introduction}

\ac{UHDR} radiotherapy is a promising cancer treatment under research, that involves an almost instantaneous delivery of a high radiation dose ($\ge$\unit{40}{\Gray}) in only a few pulses ($\leq$\unit{200}{\milli\second})~\cite{labarbe_physicochemical_2020}. The popularity of \ac{UHDR} is due to the FLASH effect, a biological effect that induces a tumour control, as in \ac{CONV-RT}, and a significant reduction of the toxicities induced on healthy tissues (skin, organs at risk)~\cite{bourhis_treatment_2019}. There is no indication that the sparing effectiveness of FLASH-RT would rest on the type of radiation but most of the pre-clinical studies investigating the FLASH effect have been conducted so far using electron beams generated by dedicated or modified clinical LINACs with energies not exceeding \unit{20}{\mega\electronvolt}~\cite{faillace_compact_2021,jorge_dosimetric_2019}. The chemical and biological mechanisms explaining the FLASH effect are still under investigation, however, a difference in term of oxygen concentration and cell reparation processes might explain the sparing effect between tumour and healthy cells~\cite{kacem_understanding_2021}.

From a clinical point of view, \ac{FLASH-RT} could allow for more effective treatment of radio-resistant tumours and would greatly reduce the number of sessions associated with a treatment, thus reducing the number of visits to the clinic, the congestion of schedules for clinicians and the overall cost of the treatment~\cite{bourhis_clinical_2019}.

Before implementation in clinics, some issues must be solved. In particular, a reliable and accurate monitoring system for measuring the dose in real time must be developed. In conventional treatments, the \ac{IC} is the reference dosimeter used in all clinical electrons and protons facilities but suffer from an upper limit on the dose rate. Above \unit{0.01}{\Gray\cdot\micro\reciprocal\second}, higher rates of the physico-chemical reactions (recombination and electronic capture) occurring in the gas induce a loss of the charge collection efficiency that leads to non-linear and non-proportional behaviors rendering them useless for beam monitoring in the case of \ac{FLASH-RT}~\cite{mcmanus_challenge_2020}. In order to resolve these issues, we propose a monitoring system that will rely on a combination of a \ac{BCT} detector for properly measuring the beam current and an \ac{IC} that will monitor the beam position and shape.

In this work, we will focus solely on the design and characterization of the \ac{BCT} and its front-end electronics in terms of noise and temporal responses.
The challenge of the detector design is to be able to monitor very low electron beam current for \ac{CONV-RT} with \unit{6\cdot 10^{-5}}{\Gray\cdot\micro\reciprocal\second} instantaneous dose rate (around \unit{10}{\micro\ampere} of beam current with a pulse of \unit{3}{\micro\second}, a pulse rate of \unit{180}{\hertz} and a mean dose rate of \unit{2}{\Gray\cdot\min^{-1}}) while maintaining a good linearity for instantaneous dose rates up to \unit{10}{\Gray\cdot\micro\reciprocal\second} (\unit{200}{\milli\ampere} within the pulse) in the case of \ac{UHDR} and a maximum surface of irradiation of \unit{78.5}{\centi\squaremetre}.

In the following, parameters that are frequency dependent have been made bold (such as $\mathbf{H}_\mathrm{HP}$) for clarity purposes.


\section{BCT requirements}
\label{sec:BCTrequirements}
\ac{BCT}s have been widely used in accelerator facilities for measuring the beam current time profile~\cite{aguilera_study_2017, stulle_beam_2019, kurita_development_2014}.
They consist of a toroidal ferrite core which conducts the magnetic field lines induced by the beam current due to its high magnetic permeability $\mu_r$, generating a current through a winding of $N$ turns of a Kapton wrapped copper wire, and a front-end electronics. 

\begin{figure}[!ht]
    \centering
    \subfigure[]{\label{fig:BCT_scheme-suba}\includegraphics[width=0.45\linewidth]{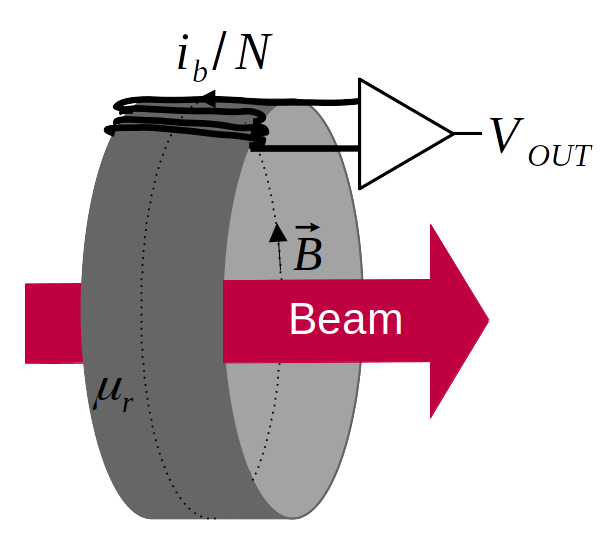}}
    \subfigure[]{\label{fig:BCT_scheme-subb}\includegraphics[width=0.45\linewidth]{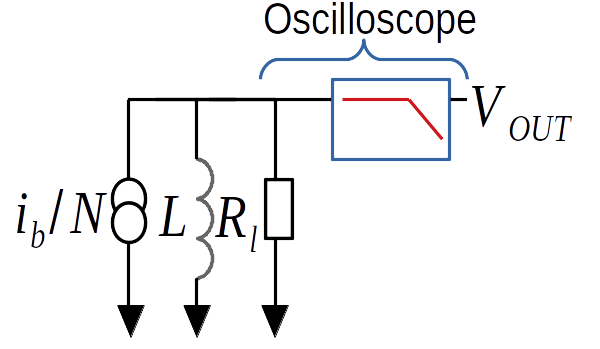}}
    \caption{(a) Schematic of the \ac{BCT} based on a toroidal ferrite core in grey, a winding of $N$ turns in black lines and a front-end electronics. (b) Simplified equivalent electronic diagram of the \ac{BCT} with $L$, the self inductance that models the ferrite core, $R_l$ the load resistor and the low pass filter of the oscilloscope used for acquisition.}
    \label{fig:BCT_scheme}
\end{figure}

Figure~\ref{fig:BCT_scheme} shows the sketch of a BCT system and its simplified equivalent electronic diagram where a load resistor $R_l$ is used to measure the current $i_b/N$ that circulates in the winding. 
 
The self inductance $L$, also represented in the diagram, that models the ferrite core is given by:
\begin{equation}
    L=A_0 \cdot N^2,
\end{equation}
where $A_0$ is the effective inductance which corresponds to the inductance of the ferrite for 1 turn of winding. $A_0$ is defined by the geometry of the ferrite and its magnetic properties. It's given by the ferrite manufacturer.

At the output of the winding, a voltage $V_{\mathrm{OUT}}$ is measured through $R_l$ using an oscilloscope having a low-pass filter. $V_{\mathrm{OUT}}$ can be evaluated as: 

\begin{equation}
    V_{\mathrm{OUT}}= \mathbf{H}_{\mathrm{LP}} \cdot \mathbf{H}_{\mathrm{HP}} \cdot  \frac{R_l}{N} \cdot i_b,
    \label{eq:Vout}
\end{equation}
where $\mathbf{H}_{\mathrm{LP}}=1/(1+\tau_{\mathrm{LP}} \cdot j\omega)$ with $\omega=2\pi f$, is the transfer function of the low pass filter of the oscilloscope with $\tau_{\mathrm{LP}}$ the rise time constant of the filter. $\mathbf{H}_{\mathrm{HP}}=\tau_{\mathrm{HP}} \cdot j\omega/(1+\tau_{\mathrm{HP}} \cdot j\omega)$ is the transfer function of the high pass filter made by the ferrite core and the winding of $N$ turns with $\tau_{\mathrm{HP}}=L/R_l=A_0 \cdot N^2/R_l$ the drop time constant of the filter.

\begin{figure}[!ht]
    \centering
    \subfigure[]{\label{fig:BCT_response-suba}\includegraphics[width=0.45\linewidth]{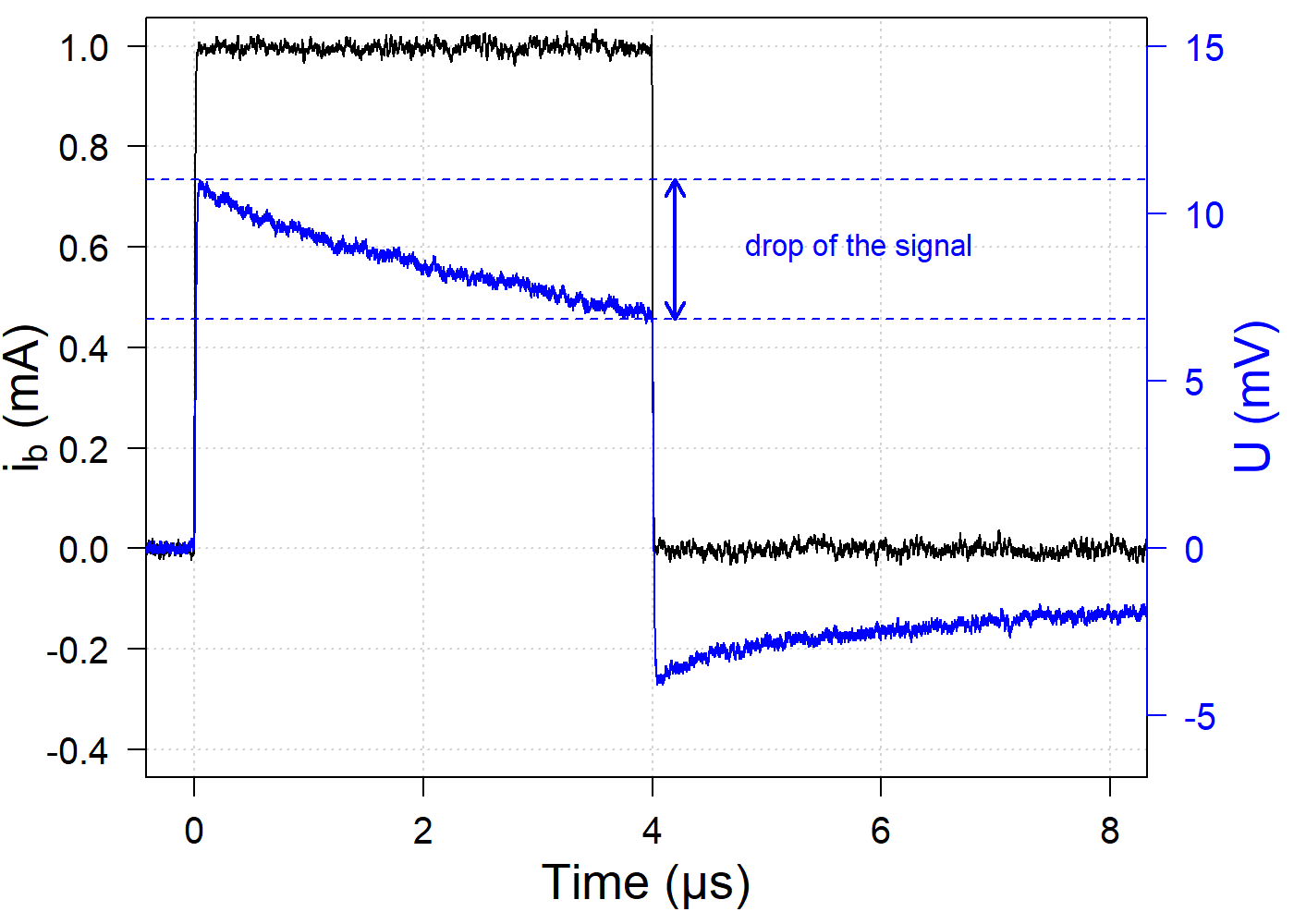}}
    \subfigure[]{\label{fig:BCT_response-subb}\includegraphics[width=0.45\linewidth]{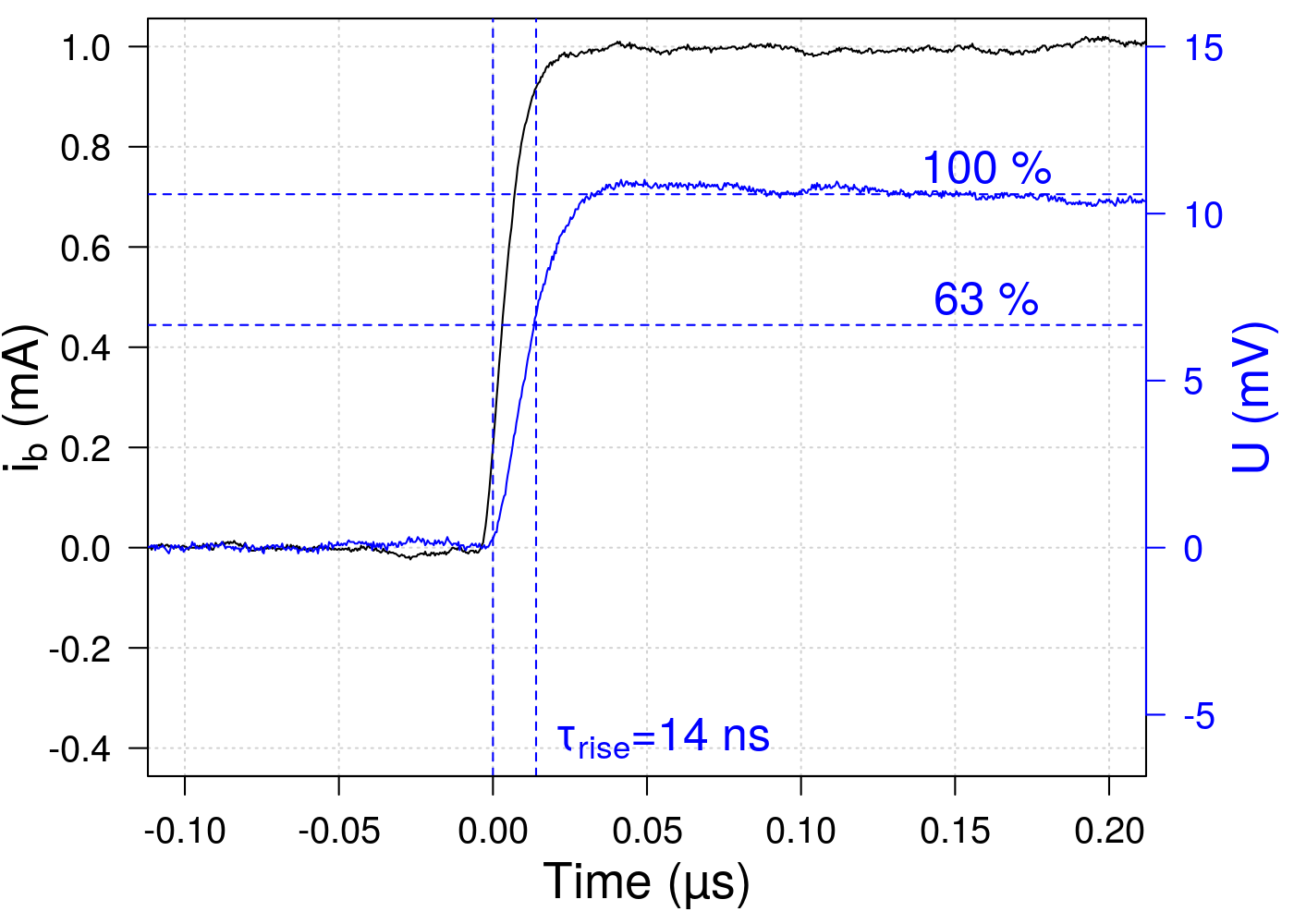}}
    \caption{The output voltage $V_{\mathrm{OUT}}$ (blue line) measured in response to a current from a pulse generator of \unit{1}{\milli\ampere} and a width of \unit{4}{\micro\second} (black line) showing (a) the drop of the signal and (b) its rise time constant $\tau_\mathrm{rise}$.}
    \label{fig:BCT_response}
\end{figure}

Figure~\ref{fig:BCT_response} shows the output signal $V_{\mathrm{OUT}}$ (blue line) measured in response to a square impulse from a pulse generator (black line) of \unit{1}{\milli\ampere} and a width of \unit{4}{\micro\second}. This measurement was made using an oscilloscope with a low-pass filter cut-off frequency of $20~$MHz.

A drop of the signal is clearly visible on the measured pulse (figure~\ref{fig:BCT_response-suba}) and defines the so-called \ac{DR} parameter as the percentage of signal lost after \unit{1}{\micro\second}. Here, the \ac{DR} value was measured at 7\%$\cdot$\textmu s$^{-1}$.
Furthermore, the rise time constant $\tau_\mathrm{rise}$ illustrated in figure~\ref{fig:BCT_response-subb} corresponds to the time after a 63\% increase in the signal from the base line. The drop time constant can also be defined as a function of the \ac{DR} by $\tau_\mathrm{HP}=63/\mathrm{DR}$ which corresponds to the time after a 63\% decrease of the signal from its maximum. These parameters characterize the time response of the system.

In order to be used for either \ac{CONV-RT} and \ac{FLASH-RT} irradiation, the \ac{BCT} system requirements are: i) $\tau_\mathrm{rise}$ must be less than \unit{20}{\nano\second} to accurately measure the beam time profile. This impacts mostly the accuracy for \ac{FLASH-RT} irradiation as the system will be used in an integrating mode in the case of \ac{CONV-RT} irradiation; ii) A current measurement relative standard uncertainty of less than 1\% to comply with the dose measurement protocol with a plane parallel ionisation chamber for electron radiotherapy~\cite{andreo_international_nodate}. To achieve this, a conservative position is to require a \ac{DR} lower than 1\% after \unit{4}{\micro\second} of beam pulse, which is the maximum pulse width value that the \ac{EF} can deliver~\cite{faillace_compact_2021}. The \ac{DR} must therefore be less than 0.25\%$\cdot${\micro\second}$^{-1}$ to avoid {\it a posteriori} numerical \ac{DR} corrections which depend on the beam pulse shape and irradiation conditions. iii) The spectral frequency band of interest of the system [$f_\mathrm{min}$;$f_\mathrm{max}$] is deduced using the drop and rise time constant by $f_\mathrm{min}=\frac{1}{2\pi\tau_\mathrm{HP}}=630$~Hz and $f_\mathrm{max}=\frac{1}{2\pi\tau_\mathrm{rise}}=8$~MHz; iv) In order to detect low beam currents, of the order of \unit{10}{\micro\ampere} in \ac{CONV-RT} irradiation, the \ac{RMS} of the beam current noise must be lower than this value.

The front-end electronic modules have been designed to meet all these requirements and are detailed in the following sections.


\section{Trans-impedance front-end electronics}
\label{sec:frontendTI}
\begin{figure}[!ht]
    \centering
    \includegraphics[width=0.35\linewidth]{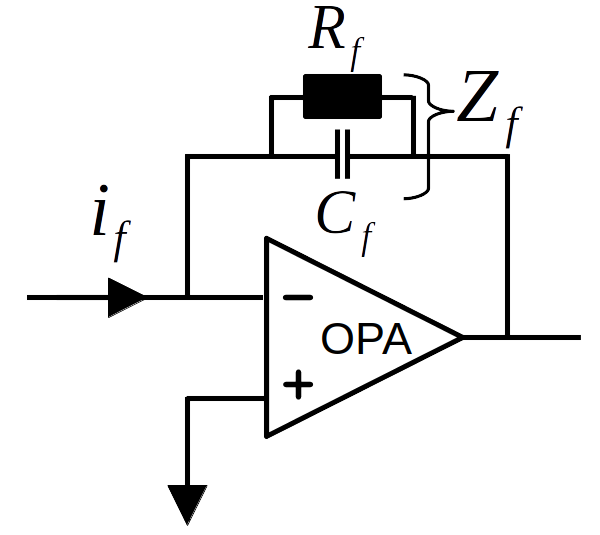}
    \caption{Equivalent electronic diagram of the trans-impedance front-end electronics with a feedback impedance $\bm{Z}_f$, an input current $i_f$ and an \ac{OPA}. }
    \label{fig:scheme_transimpedance}
\end{figure}

The solution proposed in this paper uses a front-end electronics based on a \ac{TI} circuit. Typically, a \ac{TI} circuit is composed of an \ac{OPA} and a feedback impedance $\bm{Z}_f$ which corresponds to a resistor, $R_f$, in parallel with a capacitance $C_f$, as shown in figure~\ref{fig:scheme_transimpedance}. $R_f$ defines the trans-impedance gain of the circuit. The \ac{TI} device behaves like a low-pass filter with a near-zero input impedance.

The output voltage of the \ac{TI} front-end electronics is given by:
\begin{equation}
    V_{\mathrm{OUT}}= -\mathbf{Z}_\mathrm{f}\cdot i_f,
    \label{eq:Vout}
\end{equation}
where $\mathbf{Z}_\mathrm{f}=\frac{R_f}{1+jR_f\cdot C_f\cdot \omega}$ is the feedback impedance of the \ac{TI} device with a time constant of $\tau_{\mathrm{rise}}=R_fC_f$. $i_f$ is the current that circulates in the feedback components.

In order to cover a wide range of beam currents from \unit{10}{\micro\ampere} to \unit{200}{\milli\ampere}, one \ac{TI} module per type of irradiation will be used, each having a different gain: the feedback resistor will be 10~k$\Omega$ for \ac{CONV-RT} and 100~$\Omega$ in \ac{FLASH-RT}. The choice of the feedback capacitance was imposed by the rise time constant value specified in section~\ref{sec:BCTrequirements} as:
\begin{equation}
    \tau_{\mathrm{rise}}= R_f C_f \leq 20~\mathrm{ns}.
\end{equation}

If we admit the feedback resistor values previously defined, in each mode $C_f$ is limited to:

\begin{equation}
    \mathrm{\ac{CONV-RT}: }~C_f \leq 2~\mathrm{pF~~and~~\ac{FLASH-RT}:}~C_f \leq 200~\mathrm{pF}.
\end{equation}

The components of each front-end electronics are summarized in table~\ref{tab:Feedback_components}.

\begin{table}[!ht]
    \centering
    \caption{Summary of the \ac{TI} components for each irradiation mode}
    \begin{tabular}{crr}
    \toprule
        Mode & $R_f$~({\ohm}) & $C_f$~({\pico\farad}) \\ \midrule
        \ac{FLASH-RT} & 100 & 10\\
        \ac{CONV-RT} & 10~k & 1.5\\
        \bottomrule
    \end{tabular}
    \label{tab:Feedback_components}
\end{table}

Thus, the \ac{TI} device allows to obtain the required rise time constant by selecting the suitable feedback components. Moreover, due to a near-zero input impedance, the high-pass cut-off frequency of the device is supposed to be closed to 0~Hz. The \ac{DR} being proportional to the high-pass cut-off frequency it is then supposed to be negligible. This assumption will be verified later in section~\ref{sec:optimisation}.



\section{Beam current noise definition}
\label{sec:NoiseOptim}
The \ac{RMS} beam current noise, $\sigma_b$, is a function of the noise of the front-end electronics output, $\sigma_\mathrm{OUT}$, the number of turns $N$ of the winding as well as the feedback resistor, $R_f$. 

Replacing the load resistor from figure~\ref{fig:BCT_scheme-subb} by the \ac{TI} circuit described in section~\ref{sec:frontendTI}, $\sigma_b$ can be expressed as:

\begin{equation}
    \sigma_b=\frac{N}{R_f}\cdot \sigma_\mathrm{OUT} 
    \label{eq:Ibeam}
\end{equation}

The \ac{RMS} output voltage noise $\sigma_\mathrm{OUT}$ can be computed using the \ac{PSD} of the front-end electronics, over the frequency band defined for the system from $f_{\textrm{min}}=630$~Hz to $f_{\textrm{max}}=8$~MHz, such as:

\begin{equation}
    \sigma_\mathrm{OUT}=\sqrt{\int_{f_{\mathrm{min}}}^{f_{\mathrm{max}}}\mathbf{PSD}~\mathrm{d}f}.
    \label{eq:VsRMS}
\end{equation}

For the evaluation of the \textbf{PSD}, three noise files of 1~MSa each with different sampling rate (50~kSa/s, 1~MSa/s and 100~MSa/s) were acquired at the output of the system with an oscilloscope~\cite{kg_rsrta4000_nodate} to cover the large system bandwidth.

\begin{figure}[!ht]
    \centering
    \includegraphics[width=0.6\linewidth]{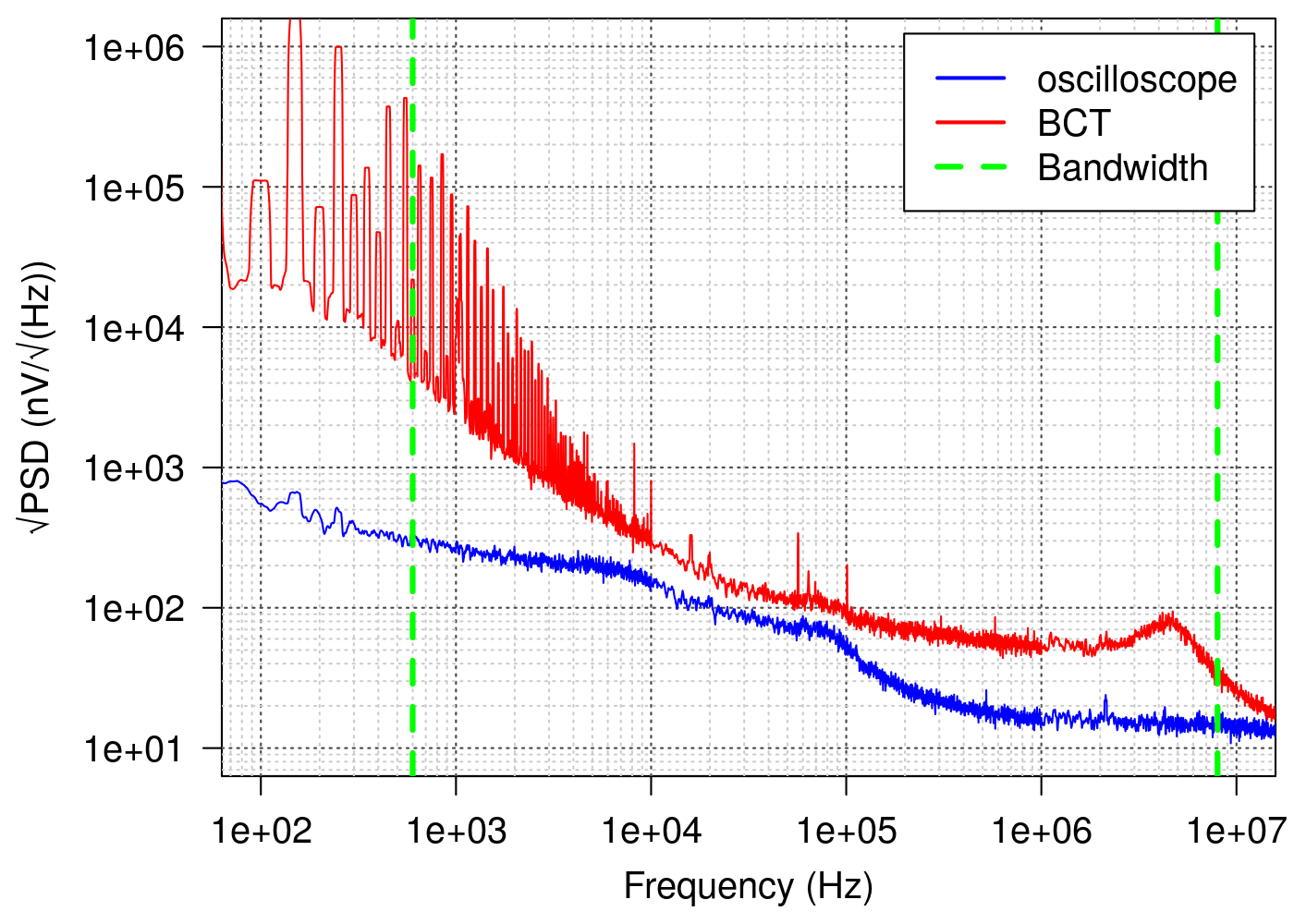}
    \caption{Square-root of measured \ac{PSD} in red for 5 turns of winding and the \ac{CONV-RT} \ac{TI} front-end electronics module used with an OPA211 amplifier. The oscilloscope \ac{PSD} is also plotted (blue line) and the required bandwidth cut-off frequencies are plotted by green dashed lines.}
    \label{fig:DSP_exemple}
\end{figure}

As an example, figure~\ref{fig:DSP_exemple} shows the square-root of the \ac{PSD} measured for 5 turns of winding in red. In comparison the square-root of \ac{PSD} of the oscilloscope is plotted in blue. For the front-end electronics, a \ac{CONV-RT} \ac{TI} module was used with an OPA211 amplifier \cite{noauthor_opa211_nodate}. The \ac{PSD} is a function of the amplifier voltage noise source, $S_{e,\mathrm{OPA}}$, the number of winding turns $N$ and the feedback resistor noise defined by $\frac{4kT}{R_f}$ which depends on the Boltzmann constant, $k$ and the temperature $T$. $S_{e,\mathrm{OPA}}$ is given by the \ac{OPA} constructor.

The voltage output noise is then computed by integrating the \ac{PSD} over the bandwidth of interest defined by the application. Finally, the output voltage noise was multipled by $N/R_f$ in order to obtain the beam current noise.


\section{Experimental set-up}

The toroidal ferrite core used for the system was chosen from the Vacuumschmelze company (reference L2160-V074 in~\cite{noauthor_amorphe_nodate}). Its aperture surface was \unit{S=120.8}{\centi\squaremetre}, chosen to be large enough for measuring beam irradiation surfaces up to $120$~{\centi\squaremetre}. For the \ac{EF} machine used for the experimental assays~\cite{faillace_compact_2021} the maximum irradiation surface is $78.5$~{\centi\squaremetre}. Its effective inductance value was $A_L=28$~\micro\henry/$N^2$ and its maximum relative permeability was 10$^5$ at frequencies up to \unit{10}{\kilo\hertz} and decreased to 2000 at 10~\mega\hertz.

\begin{figure}[!ht]
    \centering
    \subfigure[]{\label{fig:shielding-sub1} 
    \includegraphics[width=0.35\linewidth]{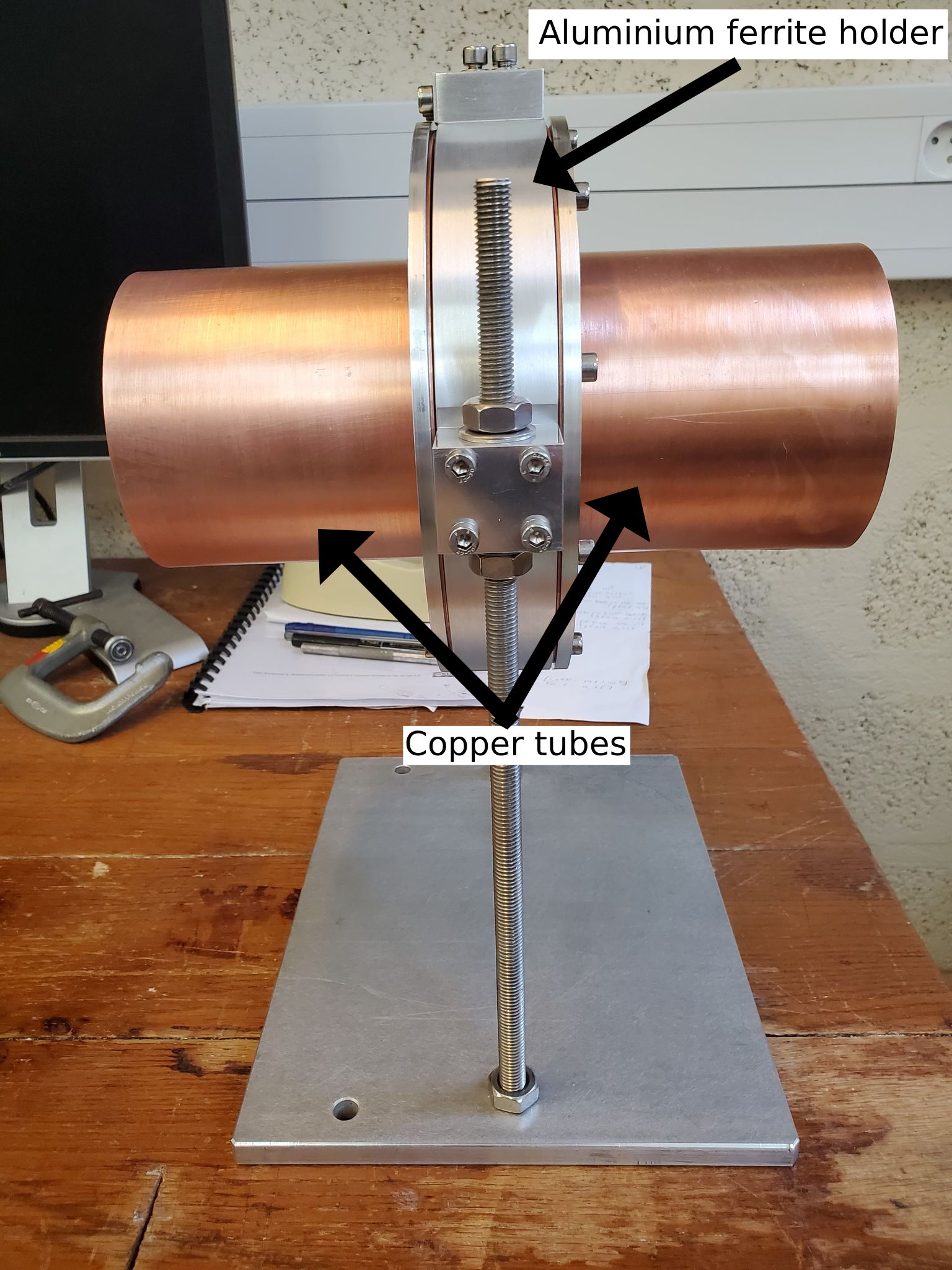}}
    \subfigure[]{\label{fig:shielding-sub2} 
    \includegraphics[width=0.35\linewidth]{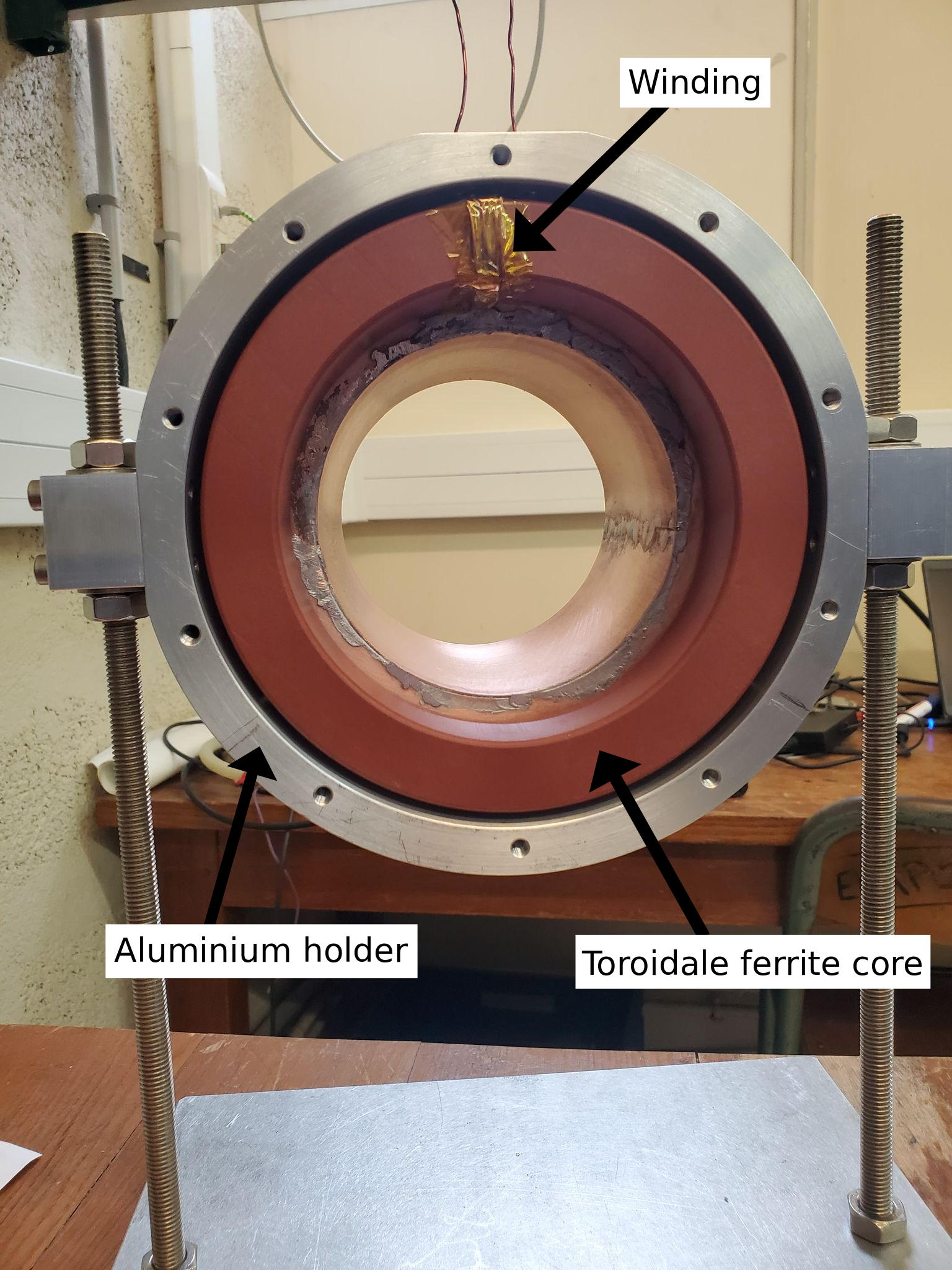}}
    \caption{Pictures of the shielding with (a) a side view of the copper tubes and the aluminium ferrite holder and (b) the toroidal ferrite core in the aluminium holder.}
    \label{fig:shielding}
\end{figure}

As a \ac{BCT} system is basically an antenna, any \ac{EM} perturbations that goes through its surface of detection result in the acquisition of extra noise sources. A shielding was then designed to reduce those perturbations to the minimum in the considered bandwidth. This shielding was made using two copper tubes of 11.5~cm length and 1.3 mm thick and an aluminium part to hold the ferrite core. Figure~\ref{fig:shielding} shows the ferrite core with its winding as well as the copper shielding and the system holder.

\begin{figure}[!ht]
    \centering
    \includegraphics[width=0.5\linewidth]{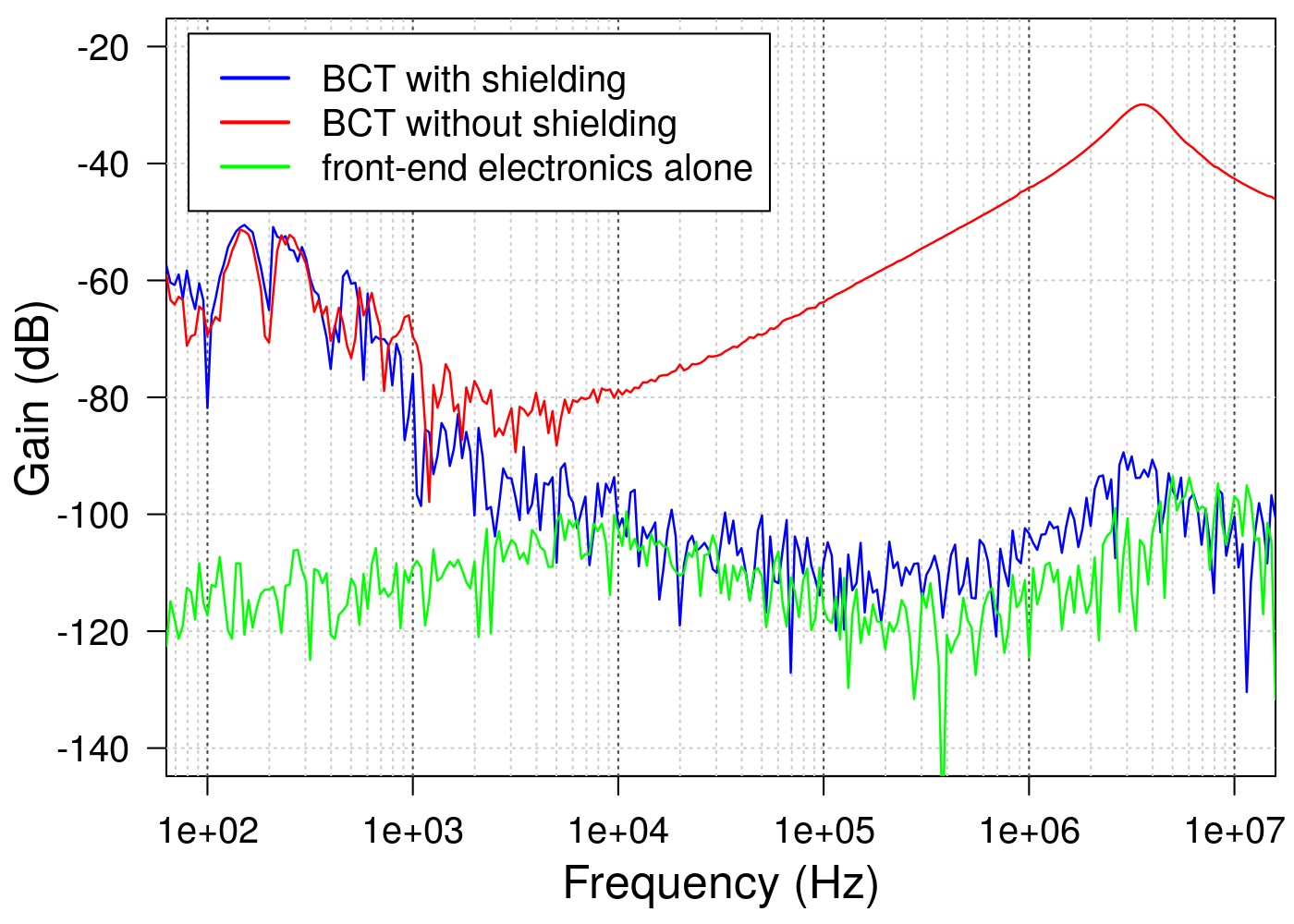}
    \caption{Measured gain of the \ac{BCT} system with and without its \ac{EM} shielding in response to a sine wave perturbation with  a current of 100~mA and a frequency varying from 100~Hz to 10~MHz. The gain of the front-end electronic alone is also plotted as reference.}
    \label{fig:Shielding_effect}
\end{figure}

The efficiency of the shielding was evaluated using a wire connected to a sine wave generator simulating \ac{EM} perturbations with a current of 100~mA and a frequency varying from 100~Hz to 10~MHz. The wire was placed outside and close to the ferrite and the gain of the \ac{BCT} system with and without the shielding was measured. Figure~\ref{fig:Shielding_effect} shows the effect of the shielding on the \ac{EM} perturbations. The shielding allows the suppression of the \ac{EM} noise above 3~kHz while the skin effect of the copper tubes has almost no impact below this frequency. The gain of the front-end electronics alone is also plotted as reference. 

\section{Optimisation}
\label{sec:optimisation}
In order to optimize the beam current noise value several amplifiers were selected for their excellent noise performances given by the constructor.

\begin{figure}[!ht]
    \centering
    \includegraphics[width=0.6\linewidth]{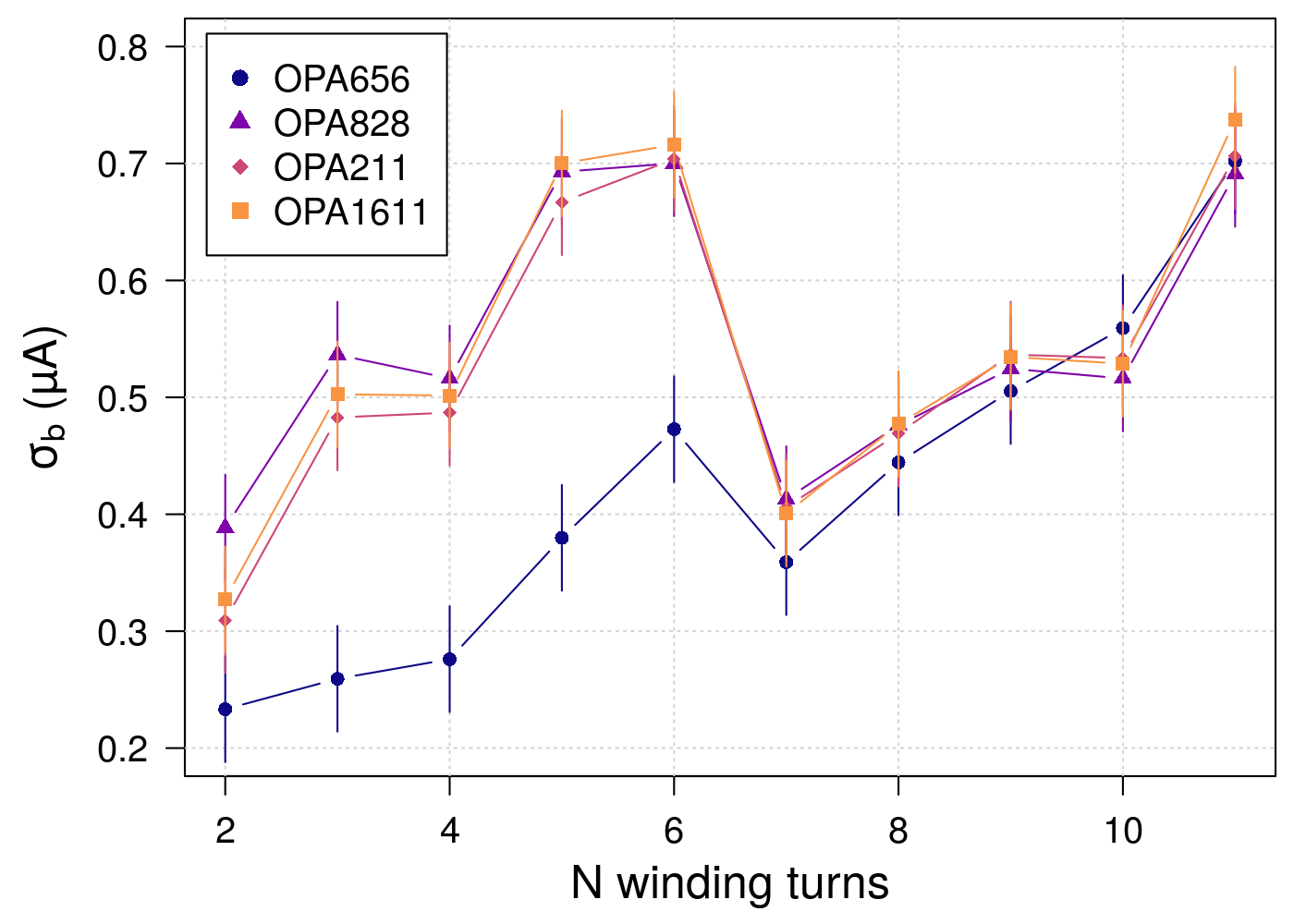}
    \caption{Experimental measurement of $\sigma_b$ for several \ac{OPA}s as a function of the number of winding turns $N$.}
    \label{fig:IRMS_optim}
\end{figure}

 An experimental evaluation of $\sigma_b$ was done for a number of winding turns varying from 2 to 11 turns and for each OPA selected in the study. Figure~\ref{fig:IRMS_optim} shows the results obtain on $\sigma_b$ for the OPA656~\cite{noauthor_opa656_nodate}, OPA828~\cite{noauthor_opa828_nodate}, OPA211~\cite{noauthor_opa211_nodate} and OPA1611~\cite{noauthor_opa1611_nodate}. For 1 turn, no signal was measured due to the magnetic field lines not being well conducted. The lowest beam current noise was measured for 2 turns and the OPA656. A local minimum was also obtained at 7 turns for any amplifier.

Moreover, the \ac{DR} of the system must also be evaluated for each \ac{OPA}s and number of turns in order to fulfill the requirement on the \ac{DR} imposed by the application. The \ac{DR} was evaluated by using the drop time constant measured with the bode diagram mode of the oscilloscope.

\begin{figure}[!ht]
    \centering
    \includegraphics[width=0.6\linewidth]{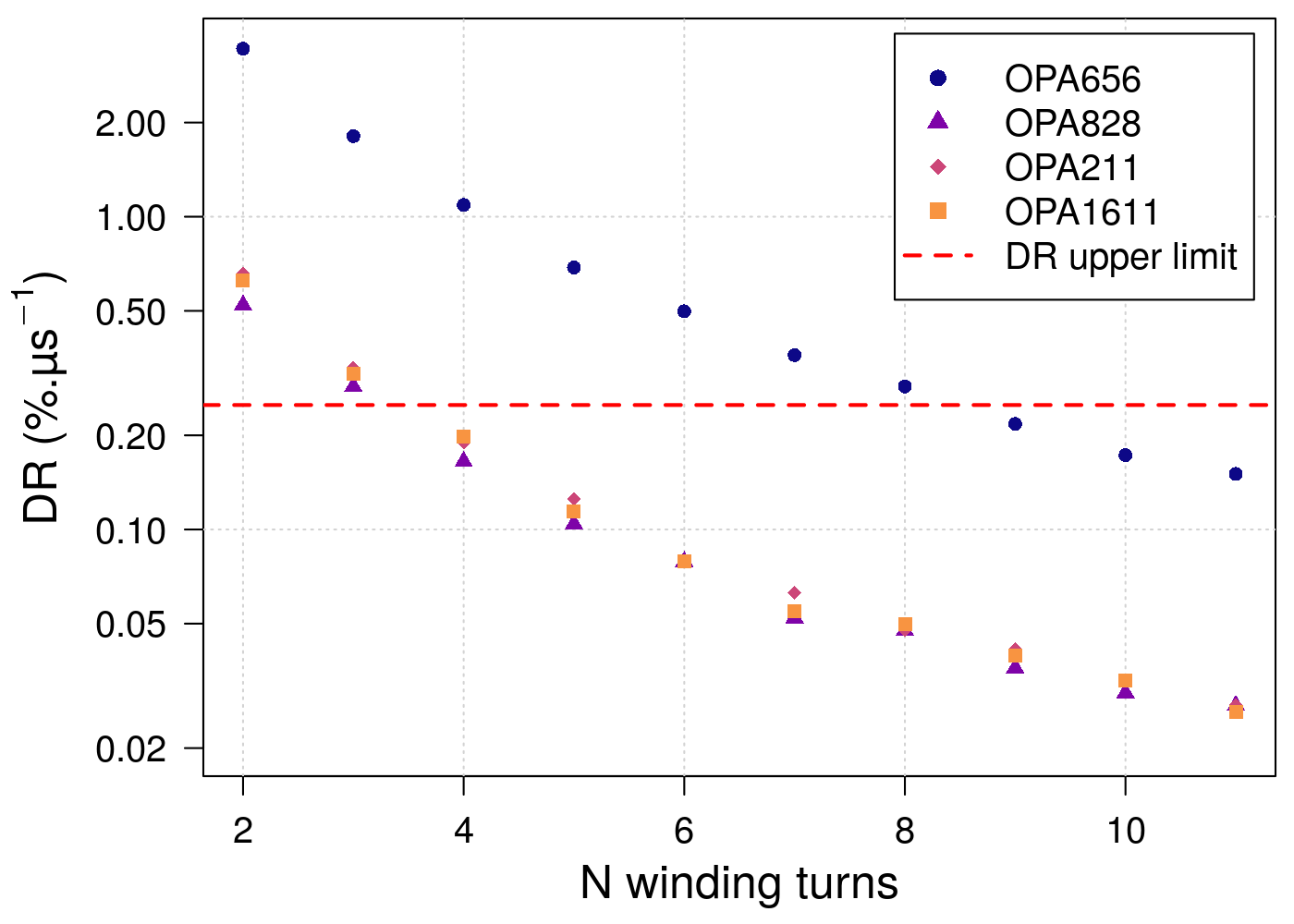}
    \caption{Experimental measurement of the \ac{DR} for several \ac{OPA}s as a function of the number of winding turns $N$.}
    \label{fig:DR_optim}
\end{figure}

 Figure~\ref{fig:DR_optim} shows the \ac{DR} in \%$\cdot$\textmu s$^{-1}$ as a function of the number of winding turns and the chosen \ac{OPA}. The upper limit imposed by the requirements is shown as an horizontal red dashed line.
 
The lowest number of turns required to achieve a \ac{DR} lower than 0.25\%$\cdot$\textmu s$^{-1}$ was 4 turns for the OPA828, OPA211 and OPA1611 and 9 turns for the OPA656. 

Using these constraints on the number of turns on the minimum beam current noise, the optimal corresponding $\sigma_{b,\mathrm{opt}}$ values obtained for the OPA828, OPA211 and OPA1611 was $(0.40\pm 0.05)$~\textmu A with an optimum number of 7 winding turns. While for the OPA656 at 9 winding turns, $\sigma_{b,\mathrm{opt}}$ was evaluated at $(0.50\pm 0.05)$~\textmu A. As OPA828, OPA211 and OPA1611 seem to have the same behavior in terms of noise response at 7~turns, the OPA828 performed slightly better considering the \ac{DR}. 

OPA828 was then chosen along with 7 winding turns. For the \ac{FLASH-RT} a \unit{50}{\ohm} resistor was added in series at the output of the circuit to divide by 2 the amplifier output current and then limit the saturation of the amplifier. The maximum output current was found around \unit{60}{\milli\ampere} for the module which corresponded to a maximum input beam current of \unit{200}{\milli\ampere}.
\newpage

\section{System performances}

\begin{figure}[!ht]
    \centering
    \subfigure[CONV-RT]{\label{fig:BCT_rise_time_response-suba}\includegraphics[width=0.48\linewidth]{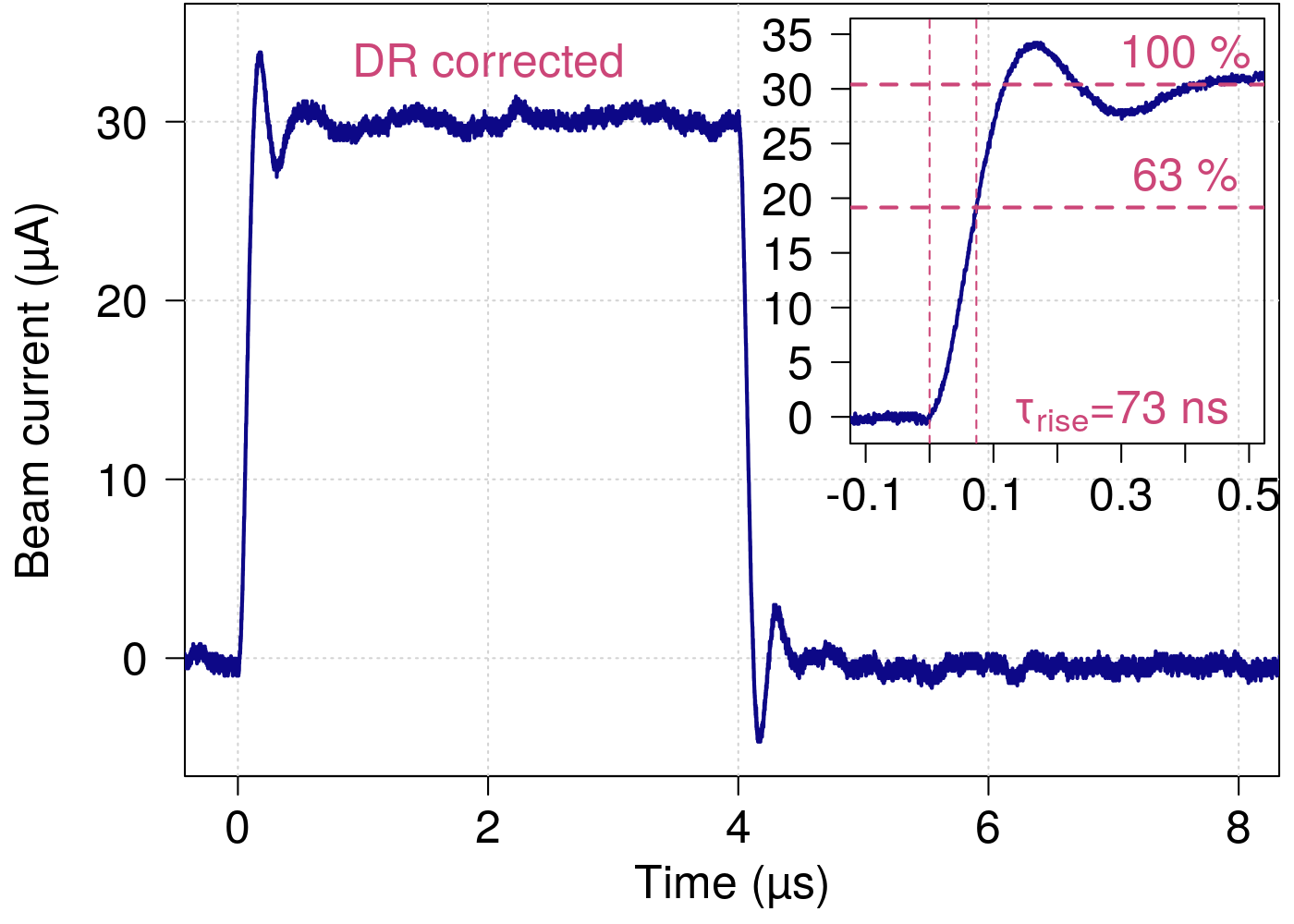}}
    \subfigure[FLASH-RT]{\label{fig:BCT_rise_time_response-subb}\includegraphics[width=0.48\linewidth]{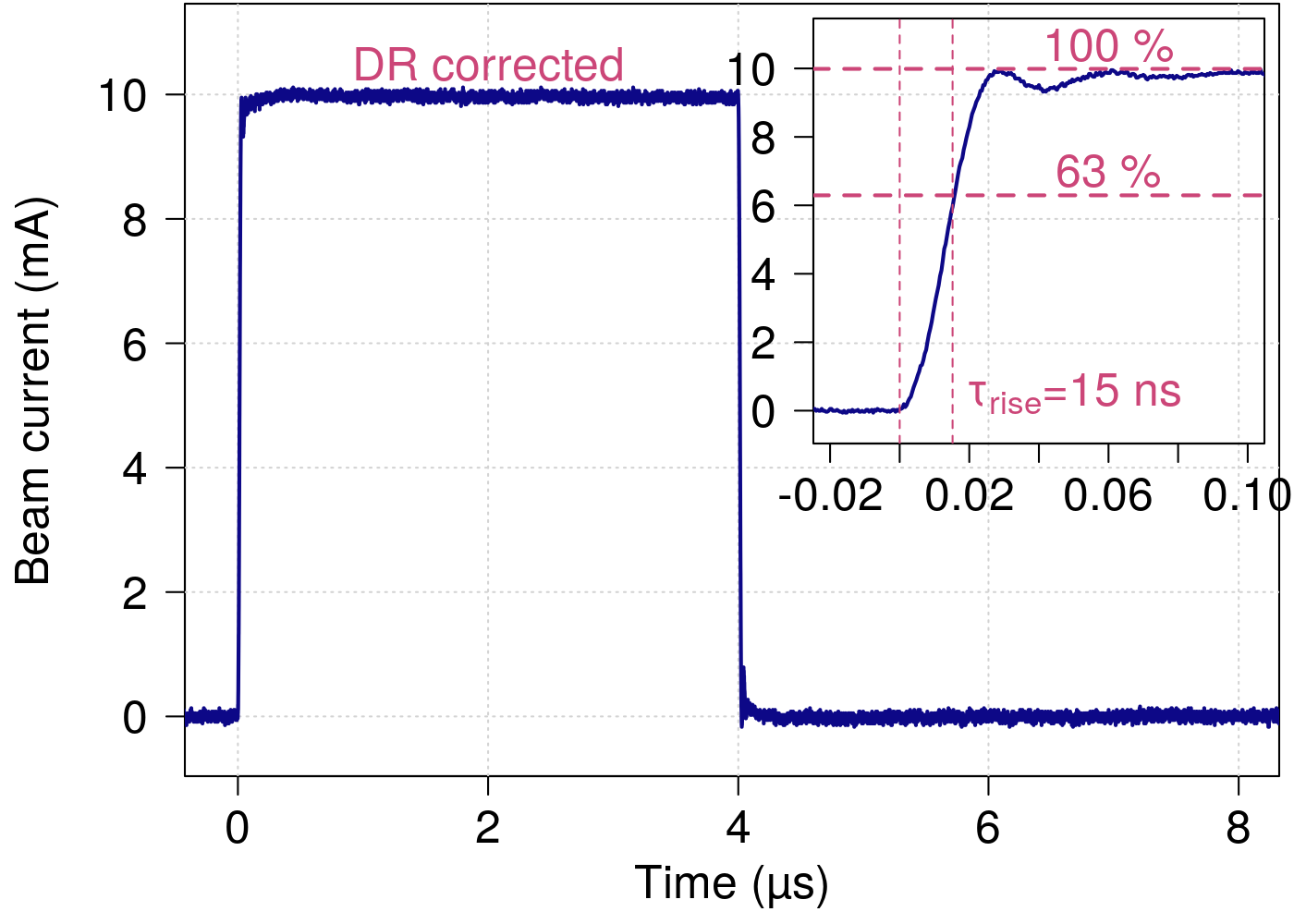}}
    \caption{Current measured by the BCT in response to a square generator pulse of (a) \unit{30}{\micro\ampere} for the \ac{CONV-RT} electronic module and (b) \unit{10}{\milli\ampere} for the \ac{FLASH-RT} showing their rise time constants $\tau_\mathrm{rise}$.}
    \label{fig:BCT_rise_time_response}
\end{figure}

The temporal responses for a square generator pulse of \unit{30}{\micro\ampere} and \unit{10}{\milli\ampere} illustrated in figures~\ref{fig:BCT_rise_time_response} show that \ac{DR} is negligible and that the rise time constant for the \ac{FLASH-RT} mode respects the requirement of a value lower than 20~ns.

The performances of the \ac{BCT} system are listed in table~\ref{tab:summary_prop} for both mode of irradiations and using an optimized number of 7 winding turns.
\begin{table}[!ht]
    \centering
    \caption{Performances of the system for both modes in terms of rise time constant, \ac{DR} and beam current noise in the frequency band [630~Hz;8~MHz], for 7 winding turns.}
    \begin{tabular}{c c c c}
         \toprule
         Mode & $\tau_\mathrm{rise}$~({\nano\second})  & \ac{DR}~(\%$\cdot${\micro\second}$^{-1}$) & $\sigma_b$~(\textmu A$_\mathrm{RMS}$)\\
         \midrule
         \ac{CONV-RT} & 73 & 0.05 & 0.40 $\pm 0.05$ \\
         \ac{FLASH-RT} & 15 & 0.4 & 3.9 $\pm 0.1$ \\
         \bottomrule
    \end{tabular}
    \label{tab:summary_prop}
\end{table}

The noise of the \ac{CONV-RT} mode module was improved by a factor of 10 compared to the \ac{FLASH-RT} \ac{TI} module.

\begin{figure}[!ht]
    \centering
    \includegraphics[width=0.7\linewidth]{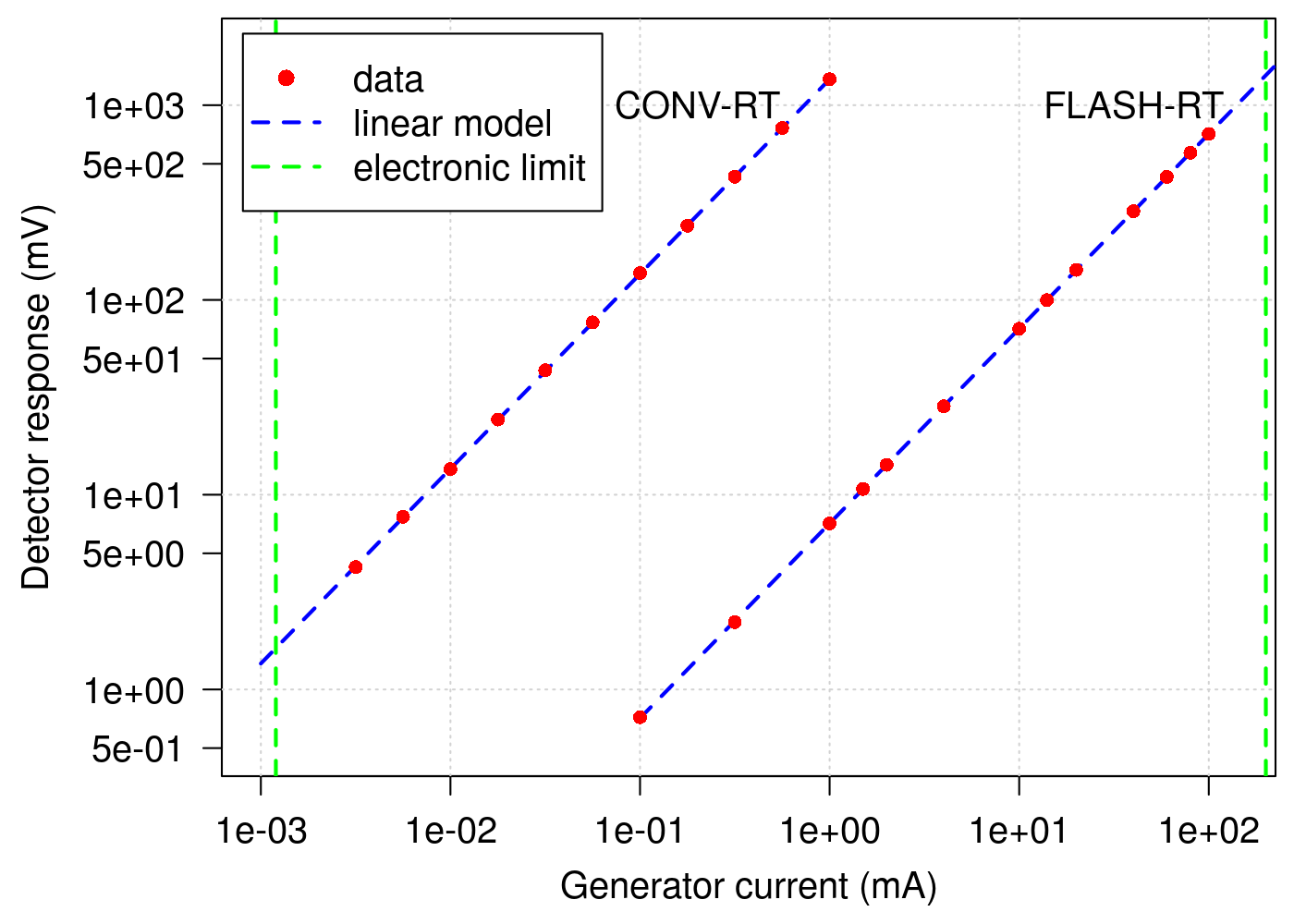}
    \caption{Calibration curves in current of the detector for the two irradiation modes.}
    \label{fig:Calibration_curve_I}
\end{figure}

Figure~\ref{fig:Calibration_curve_I} shows calibration measurements made for both modes. Data in red have been fitted using a linear model shown in blue. In addition, the electronic current limits are shown as green dashed vertical lines corresponding to the minimum and maximum beam current that the system can measure. These correspond to the noise limit of the system in \ac{CONV-RT} mode and to the saturation limit in \ac{FLASH-RT} mode. 

The calibration in current shows that the response of the detector for both modes is perfectly linear with the input current. The high beam current measurement limitation imposed at \unit{200}{\milli\ampere} is due to the amplifier saturation. For \ac{CONV-RT}, the smallest measurable signal is limited by the electronic noise level, which according to the fit and for a confidence level of 99.7\%, was established around $3\sigma_b=1.2$~\textmu A. The current measurement range of the system is [\unit{1.2}{\micro\ampere}; \unit{200}{\milli\ampere}] with a non-linearity from the fit below 1\%.
\newpage

\section{In-situ measurements for CONV-RT and FLASH-RT}

\begin{figure}[!ht]
    \centering
    \subfigure[]{\label{fig:CONV_measurement-sub1}\includegraphics[width=0.45\linewidth]{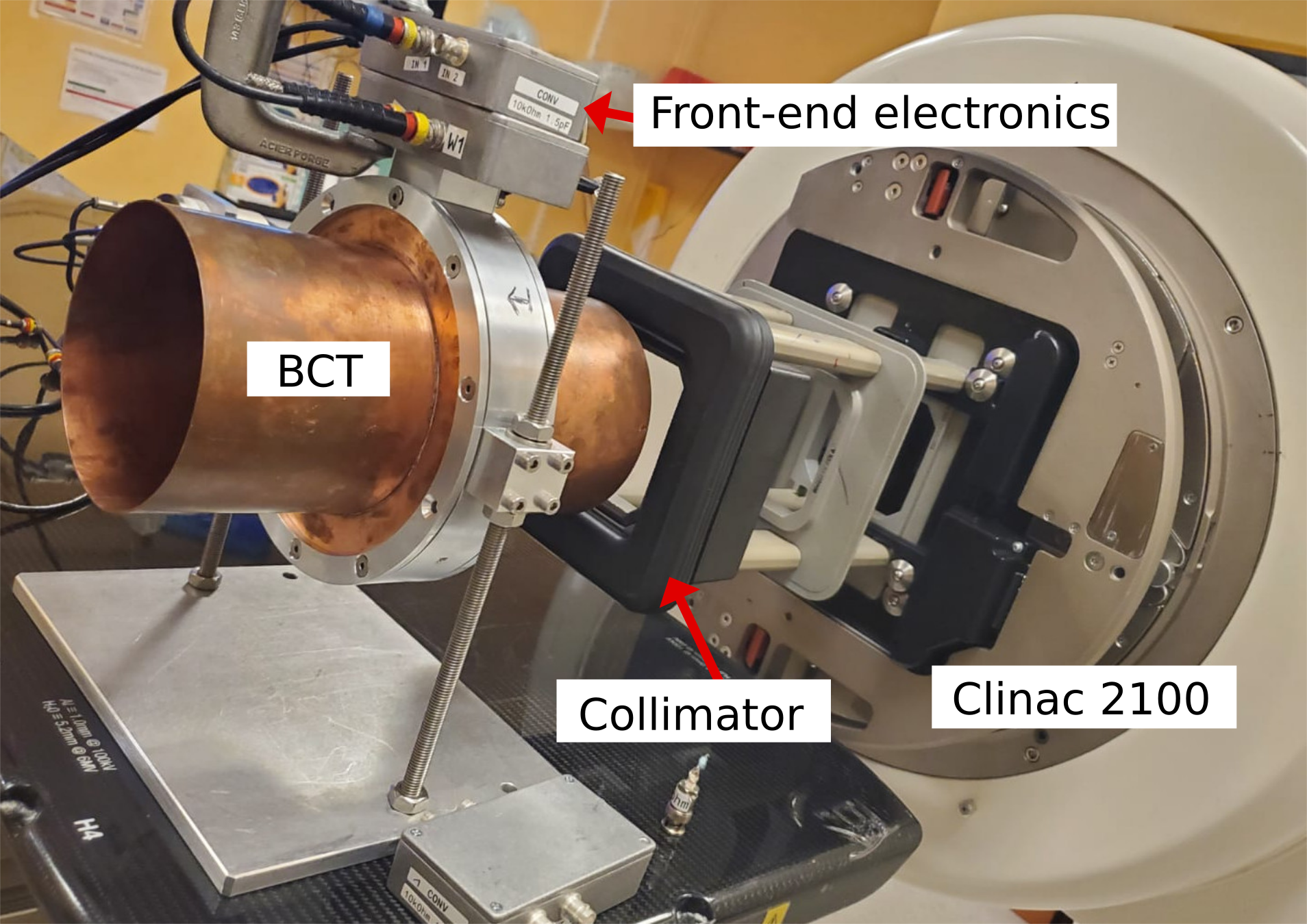}}
    \subfigure[]{\label{fig:CONV_measurement-sub2}\includegraphics[width=0.45\linewidth]{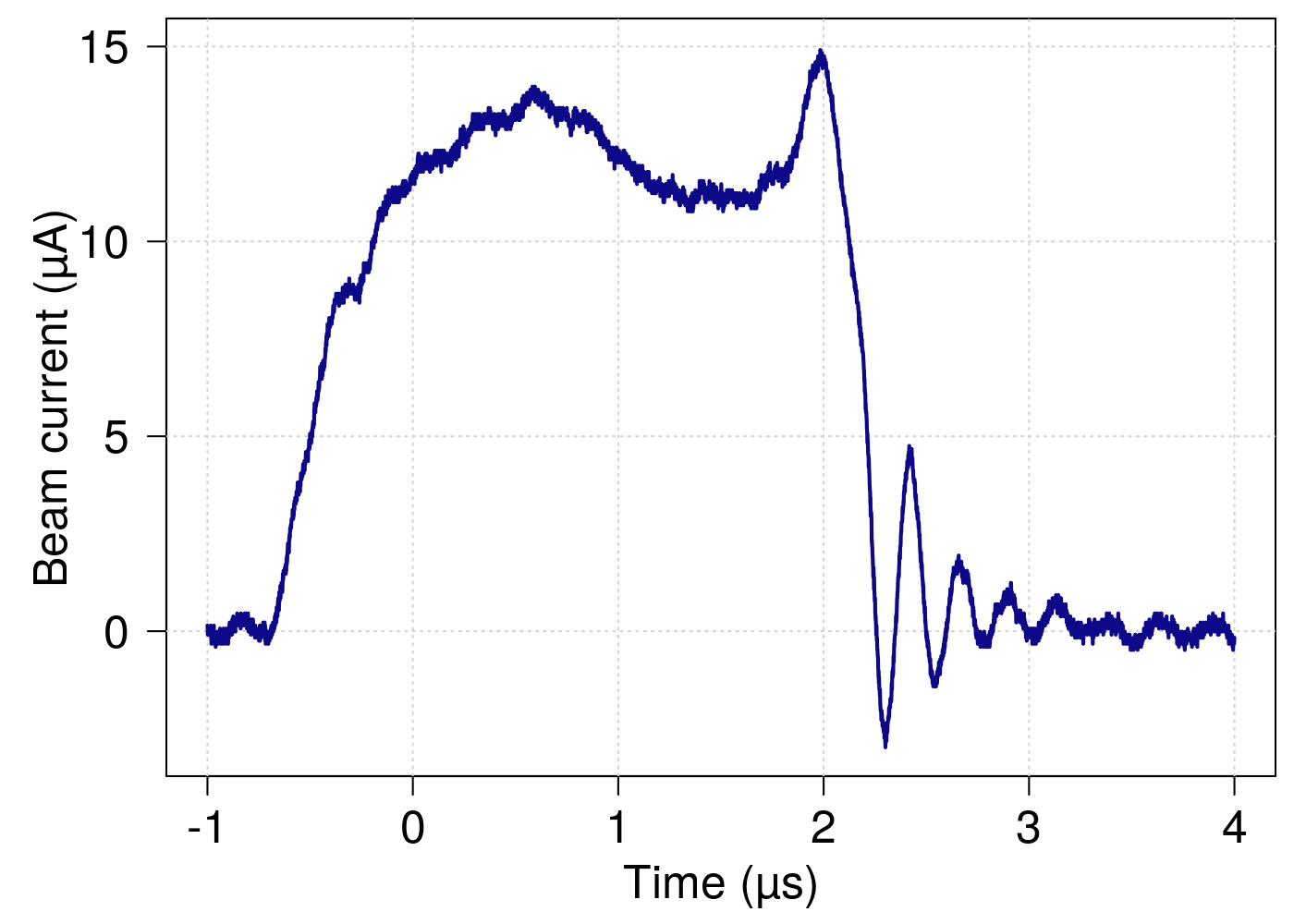}}
    \caption{Measurement performed in CONV-RT with an electron beam of 6~MeV and a mean dose rate of 2~Gy$\cdot$min$^{-1}$. a) the experimental set-up and b) an example of a pulse measured with the \ac{BCT}}
    \label{fig:CONV_measurement}
\end{figure}

Measurement have been performed in a \ac{CONV-RT} irradiation protocol using a Clinac 2100 in the François Baclesse Center (Caen, France). The machine delivered a \unit{6}{\mega\electronvolt} electron beam at 200~UM$\cdot$min$^{-1}$ corresponding to 2~Gy$\cdot$min$^{-1}$ mean dose rate during a total irradiation time of 1~min. The beam irradiation surface was 100~cm$^2$. The beam pulse width was \unit{3}{\micro\second} coupled to a repetition rate of \unit{180}{\hertz} leading to an instantaneous dose rate of \unit{3\cdot 10^{-4}}{\Gray\cdot\micro\reciprocal\second}. The \ac{BCT} was centered in the beam axis and close to the collimator. In figure~\ref{fig:CONV_measurement-sub1} a picture of the experimental set-up is shown. As an example, figure~\ref{fig:CONV_measurement-sub2} shows a pulse measured with the \ac{BCT}. The total beam charge measured by the BCT within a pulse was \unit{32.48\pm0.03}{\pico\coulomb}.

\begin{figure}[!ht]
    \centering
    \includegraphics[width=0.5\linewidth]{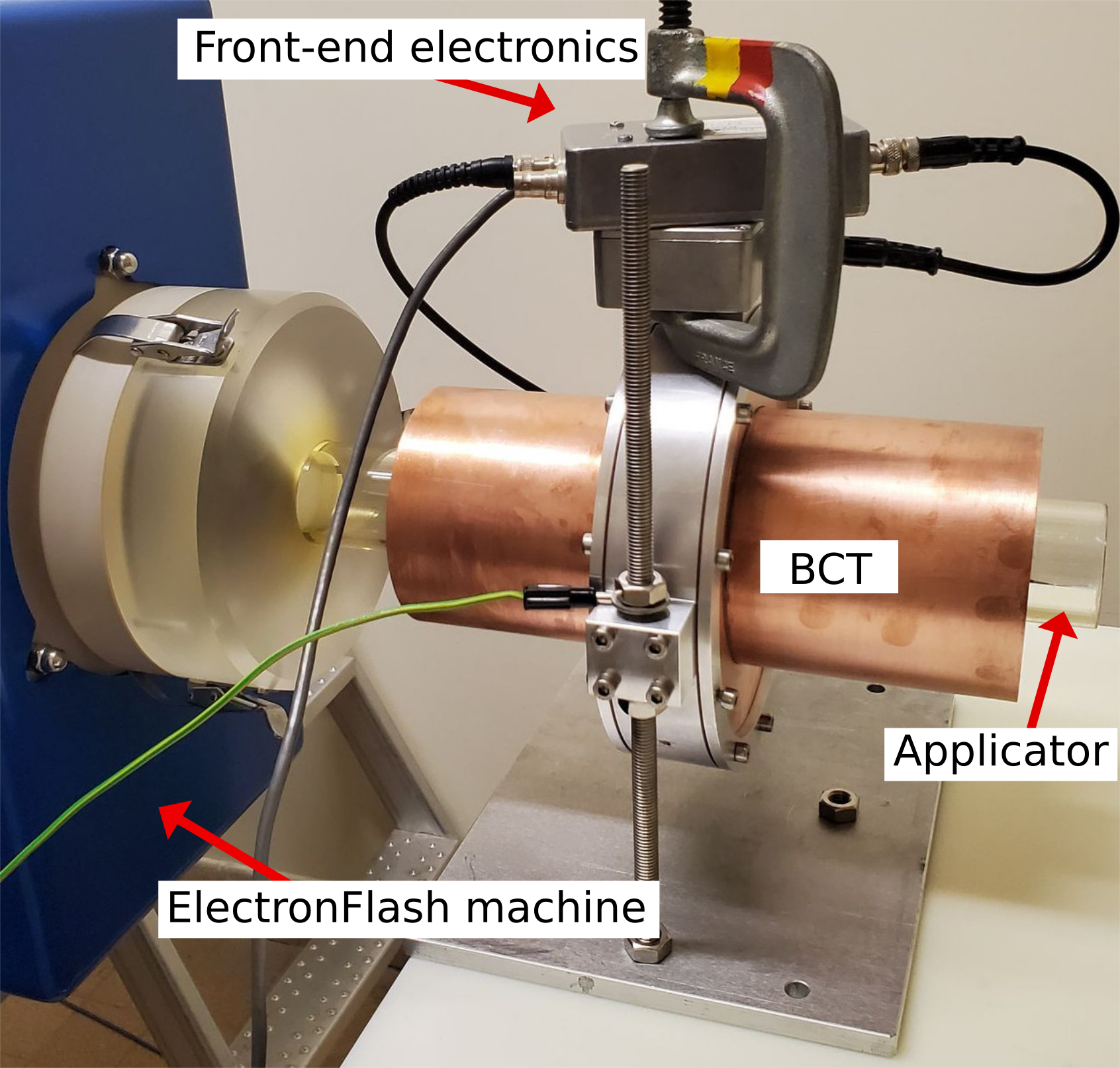}
    \caption{\ac{BCT} detector placed around the applicator used at the exit of the ElectronFlash machine}
    \label{fig:BCT_setup}
\end{figure}

The \ac{BCT} was also tested for \ac{FLASH-RT} using the \ac{EF} machine~\cite{faillace_compact_2021} located at the Curie Institute (Orsay, France). The machine delivered a pulsed electron beam of 7~MeV in \ac{FLASH-RT} mode with a instantaneous dose rate varying from \unit{5.5\cdot 10^{-1}}{\Gray\cdot\micro\reciprocal\second} to \unit{7.8}{\Gray\cdot\micro\reciprocal\second}. The dose rate was modified by changing the applicator diameter used to focus the beam to the target. Figure~\ref{fig:BCT_setup} shows a picture of the experimental set-up with ElectronFlash machine.

\begin{figure}[!ht]
    \centering
    \subfigure[]{\label{fig:FLASH_measurement-sub1} 
    \includegraphics[width=0.45\linewidth]{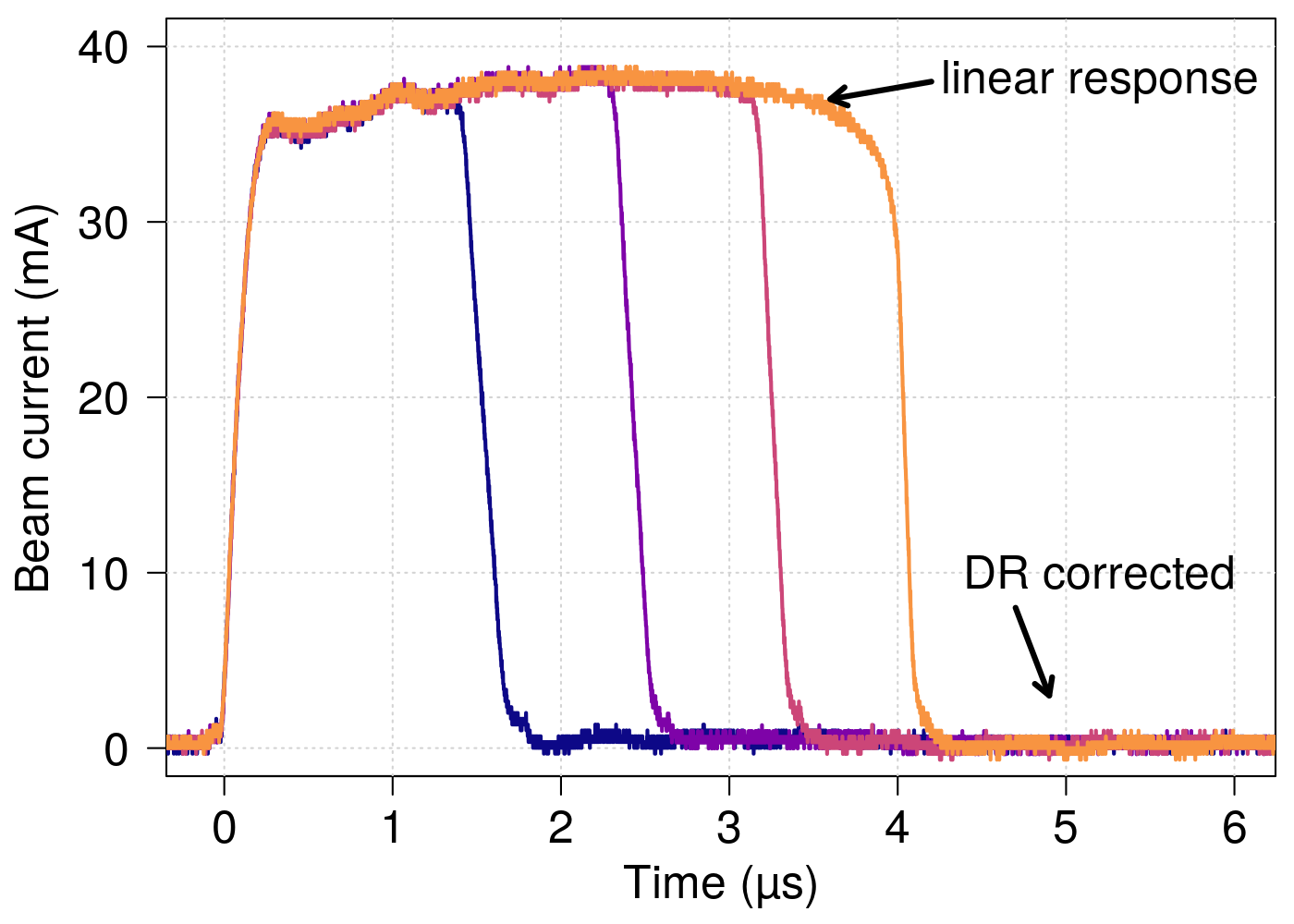}}
    \subfigure[]{\label{fig:FLASH_measurement-sub2} 
    \includegraphics[width=0.45\linewidth]{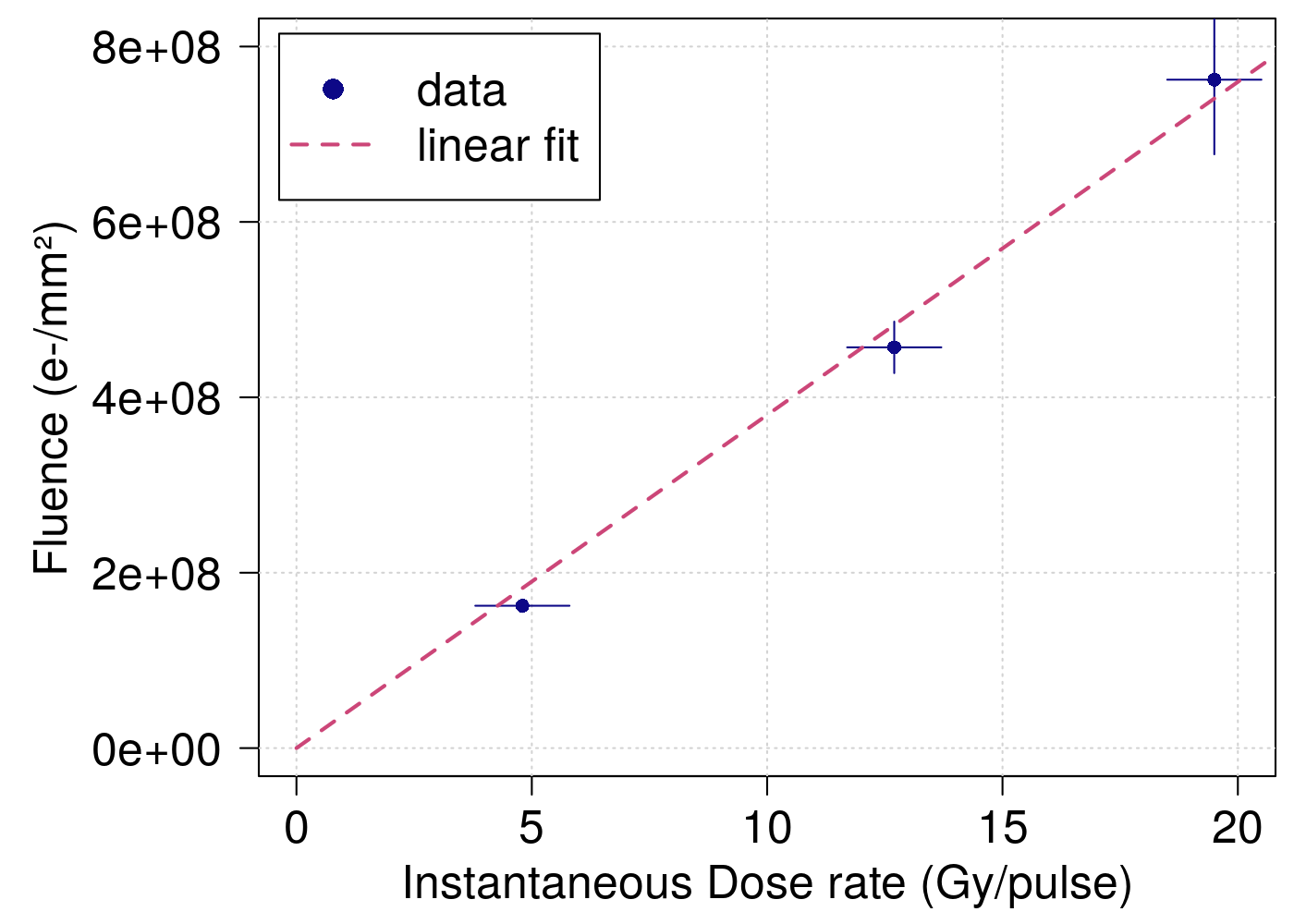}}
    \caption{Measurements performed during a FLASH-RT with an electron beam of 7~MeV for (a) several pulse widths and (b) comparison between the instantaneous dose rate measured using a gafchromic film and the fluence evaluated over 10 pulses of 4~\textmu s using the \ac{BCT}.}
    \label{fig:FLASH_measurement}
\end{figure}

Several measurements were made by changing the pulse width and the instantaneous dose rate, as shown in figure~\ref{fig:FLASH_measurement-sub1} and \ref{fig:FLASH_measurement-sub2}, respectively. 

Figure~\ref{fig:FLASH_measurement-sub1} shows that the measurement is linear with the pulse width and that the \ac{DR} is well corrected shown by the baseline recovery around 0~mA after the pulse. Moreover, we compared in figure~\ref{fig:FLASH_measurement-sub2} the instantaneous dose rate measured by a gafchromic film placed at the exit of the applicator to the beam fluence measured by the \ac{BCT}. The fluence was calculated by integrating the beam current over 10 pulses of 4~\textmu s each divided by the surface of the applicator used to deliver the beam. The uncertainty for the gafchromic film measurements was evaluated to be $\pm$1~Gy/pulse.
The linear fit applied shows a good linearity between those measurements concluding that the \ac{BCT} is almost perfectly proportional to the beam current considering the fluctuations of the ElectronFlash machine.


\section{Conclusion}
This paper proposes a complete system of Beam Current Transformer detector that provided a real time monitoring of the pulsed electron beam current from conventional to ultra-high dose rate irradiation. The used technology is based on a ferromagnetic component, which focus the magnetic field lines of the beam current. A signal is generated through a winding and processed by a trans-impedance circuit in order to amplify and correct the drop of the signal.

A new design principle is proposed in this article and an experimental optimization of the front-end electronic of the detector is performed. The prototype was tested by calibration measurements using a pulse generator to simulate the pulse beam current. Preliminary measurements were performed in \ac{CONV-RT} and \ac{FLASH-RT} electron beam by varying the pulse width and the dose rate. The fluence of the beam is measured in the beam current range [\unit{1.2}{\micro\ampere}; \unit{200}{\milli\ampere}] and the droop rate of the signal is corrected below 0.05\%$\cdot${\micro\second}$^{-1}$.

\acknowledgments
The authors would like to thank Sophie Heinrich, from the Curie Institute (Orsay, France), and Alain Batalla, from the Fran\c{c}ois Baclesse Center in  (Caen, France) for the beam access, their availability and advises.

\bibliography{main}

\end{document}